\documentclass[%
 reprint,
 amsmath,amssymb,
 aps,
pra,
 showpacs,
]{revtex4-1}

\usepackage{graphicx}
\usepackage{dcolumn}
\usepackage{bm}
\usepackage{hyperref}

\graphicspath{{converted_graphics/}}
\begin{document}

\preprint{APS/123-QED}

\title{Undamped nonequilibrium dynamics of a nondegenerate Bose gas in a 3D isotropic trap}

\author{V. E. Colussi}
\affiliation{Department of Physics, University of Colorado, Boulder, Colorado 80309-0440, USA}
\author{Cameron J. E. Straatsma}
\affiliation{JILA and Department of Electrical, Computer, and Energy Engineering, University of Colorado, Boulder, CO 80309-0440}
\author{Dana Z. Anderson, M. J. Holland}
\affiliation{Department of Physics and JILA, University of Colorado, Boulder, CO 80309-0440, USA}

\date{\today}

\begin{abstract}
We investigate anomalous damping of the monopole mode of a non-degenerate 3D Bose gas under isotropic harmonic confinement as recently reported by the JILA TOP trap experiment [D. S. Lobser, A. E. S. Barentine, E. A. Cornell, and H. J. Lewandowski (in preparation)]. Given a realistic confining potential, we develop a model for studying collective modes that includes the effects of anharmonic corrections to a harmonic potential. By studying the influence of these trap anharmonicities throughout a range of temperatures and collisional regimes, we find that the damping is caused by the joint mechanisms of dephasing and collisional relaxation. Furthermore, the model is complimented by Monte Carlo simulations which are in fair agreement with data from the JILA experiment.
\end{abstract}

\maketitle

\section{\label{sec:level1}Introduction\protect\\}
In the late 19th century, Maxwell and Boltzmann uncovered a path to connect the Newtonian mechanics of molecular dynamics to the hydrodynamic equations of Euler and Navier-Stokes.  These kinetic theories, of which the fluid theories were limiting cases, established a platform to formulate and analyze the Maxwell distribution and Demon \cite{Max1,Max2}, Boltzmann's equation and the H-theorem, and the assumption of molecular chaos \cite{Bolt} (Stosszahlansatz).  These fundamental ideas, in particular the H-theorem, famously stirred controversy in the scientific community, which in the 1890s was centered around the  ``reversal" paradox and the ``recurrence" paradox based on Poincar\'e's theorem \cite{Br2}.   Lesser known, but equally as curious was the discovery by Boltzmann of a class of exact solutions to his kinetic theory of which the Maxwell distribution is a special case \cite{boltzmann}.  Such solutions include the undamped, nonequilibrium oscillations of a classical gas under 3D isotropic harmonic confinement:  the so-called breathing or monopole mode.  Whereas some of the unexplored aspects in the justification of the Boltzmann equation began to be filled in the early 20th century, experimental demonstration of the undamped monopole mode oscillation has yet to be fully investigated.    

Advances in the trapping of ultracold gases have allowed for the study of collective modes under harmonic confinement.  Indeed, many experiments have probed the transition between the collisionless and hydrodynamic regimes \cite{8,9,10,12,13,14,15,16,17,18,19,20,21,22,23} in anisotropic traps.  Such experiments measure the eigenfrequencies and damping rates of the various multipole modes and are generally limited by three-body losses approaching the hydrodynamic limit.  Engineering an isotropic 3D trap of sufficiently perfect symmetry has remained a technological hurdle in investigating the undamped oscillation of the monopole mode.   

A recent experiment from the group of Eric Cornell at JILA \cite{dan} involving a nondegenerate cloud of $^{87}$Rb well above the transition temperature utilized a TOP trap with anisotropies as low as $0.02\%$.  The monopole mode was made to oscillate with minimal damping over many trap periods by dithering the trap frequency.  However, the damping observed was anomalously large given the level of anisotropy present in the system.  

This paper explores the inherent limitations on the observation of an undamped, nonequilibrium monopole mode given a realistic trapping scenario, using the JILA experiment as illustration.  The effects of slight anisotropies in the trap are first quantified and shown to lead to coupling between the monopole mode and the quadrupole modes.  The quadrupole modes suffer from collisional damping in the transition between hydrodynamic and collisionless limits.  The importance of cubic and quartic anharmonic corrections to the trapping potential are then considered.  Such corrections give rise to a temperature dependent eigenfrequency and damping rate for the collective modes.  They also lead to dephasing effects that create the appearance of actual decoherence in the system.  In general, such effects generate additional features observed in the damping in the transition between collisionless and hydrodynamic limits.  This in turn provides an explanation for the anomalous damping seen in the JILA experiment.  

\section{\label{sec:level2}Background}

\subsection{Boltzmann Equation }
The canonical Boltzmann equation \cite{huang,baym} describes the phase space evolution of a cloud of thermal atoms obeying classical statistics:
\begin{equation}
\frac{\partial f({\bf r},{\bf v_1},t)}{\partial t} +\{f({\bf r},{\bf v_1},t),H \}=I_{coll}[f]\label{boltz},
\end{equation}
where  $f({\bf r},{\bf v_1},t)$ is the single particle phase space distribution that depends on the generalized coordinate ${\bf r}$ and velocity ${\bf v_1}$ of all the particles, $H$ is the single particle Hamiltonian, and $\{\}$ denotes the classical Poisson bracket. The left-hand side describes the single-particle evolution of the thermal cloud and the right-hand side is the collision integral which drives the system toward a state of statistical equilibrium  
\begin{widetext}
\begin{equation}
I_{coll}[f]=\frac{\sigma}{4\pi}\int d^2\Omega d^3v_2|{\bf{v_2}}-{\bf{v_1}}|[f({\bf{v_1}}')f({\bf{v_2}}')-f({\bf{v_1}})f({\bf{v_2}})]\label{Icoll}
\end{equation}
\end{widetext}
at a rate determined by the total scattering cross section $\sigma$.  The velocities of the incoming particles are ${\bf v_1}$ and ${\bf v_2}$, and the outgoing particles leave with velocities ${\bf v_1'}$ and ${\bf v_2'}$.  For binary elastic collisions, both the center of mass (COM) momentum and the magnitude of the relative momentum are conserved, and the solid angle integral is over the full range of possible final directions for the relative momentum.  

As a direct consequence of Boltzmann's H-theorem, the necessary condition for the vanishing of the collision integral is that $\log(f)$ be composed of quantities that are elements in the set of collisional invariants $\{\chi_1,\chi_2,...\}$.  This defines a class of solutions of the form 
\begin{equation}
\log(f({\bf r},{\bf v},t))=\chi_1+\chi_2+...
\end{equation}
For a binary elastic collision, this set is spanned by the kinetic energy, momentum, and any velocity independent function, giving
\begin{equation}
\log\left (f({\bf r},{\bf v},t)\right)=a+{\bf b}\cdot {\bf v}+c{\bf v^2},\label{qcons}
\end{equation}
 where $a,\bf{b},c$ are possibly functions of position and$/$or time.  The familiar Maxwell-Boltzmann distribution of a stationary dilute gas follows from this expression for the choice $|{\bf b}|=0$, $c=-1/k_b T$ and $a=\log n\left(m/2\pi k_b T\right)^{3/2}$.

Although the parameters that lead to the Maxwell-Boltzmann distribution are the most  well-known; it should be emphasized that Boltzmann noted a curious class of parameters that corresponded to collective motion in a harmonic trap \cite{boltzmann}.  Boltzmann considered solutions where $c$ had periodic time-dependence describing temperature oscillations, as well as a position and time-dependent ${\bf b}$.  This class of solutions corresponds to the so-called breathing or monopole mode and describes undamped collective motion of a thermal cloud in an isotropic 3D trap, regardless of the oscillation amplitude and interaction strength.  In a recent paper \cite{gueryNE}, the class of undamped nonequilibrium solutions permitted by the H-Theorem was extended to time-dependent confinement.  For the remainder of the paper only the case of harmonic confinement is discussed.

\subsubsection{Collective modes in a harmonic trap.}

The collective modes of a 3D isotropic trap with trapping frequency $\omega_0$ reflect the underlying spherical symmetry and the quadratic form of the trapping potential.  Any mode of the trap can be expressed in terms of irreducible spherical tensors, in this case they are multipole modes denoted by the labels $(l,m)$.  The $l=0$ mode is the monopole mode of the system, and is characterized by an undamped oscillation at $2\omega_0$.  The $l=1$ dipole mode is a COM oscillation independent of the character of the interaction and therefore undamped.  The $l=2$ quadrupole mode oscillates at $2\omega_0$ in the collisionless regime and $\sqrt{2}\omega_0$ in the hydrodynamic regime.  It is collisionally damped in the hydrodynamic crossover between the two extremes.  For small amplitude oscillations about equilibrium, one finds a general form for the oscillation frequency $\omega=\omega_r+i\Gamma$ of the collective mode \cite{guery1999}:
\begin{equation}
\omega^2=\omega^2_{CL}+\frac{\omega^2_{HD}-\omega^2_{CL}}{1+i\omega\tilde{\tau}_{coll}},\label{relaxation}
\end{equation}
where the subscripts denote the mode frequency in the hydrodynamic (HD) and collisionless (CL) limits, and $\tilde{\tau}_{coll}$ is related to $\tau_{coll}$, the thermal relaxation time, by $\tilde{\tau}_{coll}=(\omega^2_{CL}/\omega^2_{HD})\tau_{coll}$.  The imaginary part of $\omega$ corresponds to the damping and the real part to the mode frequency.  This general form is common to many kinds of relaxation phenomenon \cite{landau}, and, assuming $\Gamma\ll\omega_r$ can be written in an asymptotic form near the collisional and hydrodynamic regimes:
\begin{eqnarray}
\Gamma_{HD}&\approx&\frac{\tilde{\tau}_{coll}}{2}(\omega_{CL}^2-\omega_{HD}^2),\nonumber\\
\Gamma_{CL}&\approx&\frac{1}{2\omega^2_{CL}\tilde{\tau}_{coll}}(\omega_{CL}^2-\omega_{HD}^2).\label{dampinglimit}
\end{eqnarray}
In the transition regime, again provided that $\Gamma\ll\omega_r$, the damping takes on the simple form \cite{landau}:
\begin{equation}
\Gamma\approx \frac{\tilde{\tau}_{coll}}{2}\frac{\omega_{CL}^2-\omega_{HD}^2}{1+\omega_{CL}^2\tilde{\tau}_{coll}^2}.\label{transitionlimit}
\end{equation}

For traps where the spherical symmetry is broken, the monopole mode couples to the quadrupole modes.  It is therefore collisionally damped at the hydrodynamic crossover occurring between the hydrodynamic and collisionless regimes.  In Sec.~III, the impact and strength of this coupling on the monopole mode is discussed.  

\subsection{JILA TOP trap experiment $^{87}$Rb}

Of the many schemes for magnetically trapping ultracold neutral atoms, the TOP trap \cite{TOP} allows flexibility in tailoring the level of anisotropy present in the system.   TOP traps utilize a fast rotating bias field oriented perpendicular to a quadrupole magnetic field.  This avoids Majorana losses that would otherwise occur due to the zero point of the field being inside the cloud~\cite{Majorana}.  The quadrupole and fast rotating bias fields produce a magnetic field that has Cartesian components
\begin{equation}
\vec{B}=\left(
\begin{array}{c}
\frac{B_z}{2}x+B_0\cos(\Omega t)\\
\frac{B_z}{2}y+B_0\sin(\Omega t)\\
-B_z z
\end{array}
\right)\label{TOPbfield}
\end{equation}
 where $B_0$ is the field strength of the bias field, $B_z$ is the strength of the quadrupole field.  This bias field rotates at angular frequency $\Omega$, which is typically much larger than the Larmor frequency.  Near the minimum of the trap the cooled atoms experience a time averaged potential 
 \begin{equation}
 U_{\rm TOP}(r,z)=\mu B_0+\frac{\mu B_z}{16 B_0}(r^2+8z^2)\label{Utop}
 \end{equation}
 where $\mu$ is the magnetic moment and $r=\sqrt{x^2+y^2}$.  
 
 In the high-field limit $(\mu B_z\gg mg)$ the trap is anisotropic with aspect ratio $\omega_z/\omega_r=\sqrt{8}$.  As the quadrupole field strength is decreased, the effects of gravity become important as the atoms in the trap sag away from the high-field limit trap minimum.  The extent of the sag is characterized by the dimensionless quantity $\Lambda=mg/\mu B_z$ and shifted trap minimum $z_0=-\left(B_0\Lambda\right)/\left(B_z\sqrt{1-\Lambda^2}\right)$.  The potential expanded about the sag position $z_0$ is
\begin{eqnarray}
U_{\rm sag}(r,z)=&\mu B_0\sqrt{1-\Lambda^2}+\frac{\mu B_z^2}{16 B_0}(1+\Lambda^2)r^2\sqrt{1-\Lambda^2}\nonumber\\
&+\frac{\mu B_z^2}{2B_0}z^2(1-\Lambda^2)^{3/2}.\label{sag}
\end{eqnarray}
The ratio of the trap frequencies is
\begin{equation}
\frac{\omega_z}{\omega_r}=\sqrt{8\frac{1-\Lambda^2}{1+\Lambda^2}}\label{sagfrequencies}
\end{equation}
with $\Lambda=7/9$ giving an isotropic trapping potential.   For further details of the modified TOP trap used in the JILA experiment, we refer the reader to Ref.~{\cite{DanThesis}}.  

Utilizing gravity to symmetrize the trap results in trapping frequencies on the order of 10 Hz.  In such a loose trap anharmonic corrections become important and must be included in calculations.  The potential used in calculations involving the anharmonic corrections has the form
\begin{eqnarray}
U(r,z)_{\rm ah}=&\frac{m}{2}\left(\omega_z^2 z^2+\omega_r^2 r^2\right)+\frac{m\alpha}{3} z^3+\frac{m\beta}{3}r^2z\nonumber\\
&+\frac{m\kappa}{4}z^4+\frac{m\delta}{4}r^2z^2+\frac{m\epsilon}{4}r^4\label{potential}
\end{eqnarray}
for general coefficients of the cubic and quartic terms (see Ref.~{\cite{DanThesis}}).

The monopole mode is driven experimentally by applying a sinusoidal variation in the trapping frequency over four periods of monopole oscillation.  The strength of the TOP field is changed along with a vertical bias which fixes the minimum of the potential (see Ref.~{\cite{DanThesis}}).  The net result is a roughly $25\%$ increase in the mean cloud size.

\section{\label{sec:level3}Theory}
In this Section, two approaches to modeling small amplitude collective excitations of a nondegenerate Bose gas about its equilibrium state are discussed.
\subsection{Moment method}
By taking moments of the Boltzmann equation Eq.~(\ref{boltz}), a set of coupled equations of motion can be derived that describe the dynamical evolution of various collective modes \cite{huang}.  For an arbitrary quantity $\chi({\bf r},{\bf v})$, the equation of motion is 
\begin{equation} 
\frac{d\langle \chi \rangle}{dt}-\langle {\bf v}\cdot \nabla_{\bf r}\rangle-\left\langle\frac{{\bf F}}{m}\cdot\nabla_{\bf v}\chi\right\rangle=\langle \chi I_{coll}\rangle ,\label{momenteq}
\end{equation}
where the average quantity is given by
\begin{equation}
\langle \chi \rangle=\frac{1}{N}\int d^3 r d^3 vf({\bf r},{\bf v},t)\chi({\bf r},{\bf v}),\label{avgmoment}
\end{equation}
and the collisional average of this quantity is given by 
\begin{equation}
\langle \chi I_{coll}\rangle= \frac{1}{4N}\int d^3 r d^3 {\bf v_1} \Delta \chi I_{coll}[f],\label{avgcoll}
\end{equation}
where $\Delta \chi=\chi_1+\chi_2-\chi_{1'}-\chi_{2'}$ with $\chi_i=\chi({\bf r},{\bf v_i})$.  

From Eq.~(\ref{momenteq}), it is straightforward to form a closed set of coupled equations for the quadratic moments, recalling that moments of the form in Eq.~(\ref{qcons}) have no collisional contribution.  For an isotropic trap, taking the moment $\langle {\bf r}^2 \rangle$ yields a closed set of equations describing the evolution of the monopole mode:
\begin{eqnarray}
\frac{d \langle {\bf r}^2\rangle}{dt}-2\langle {\bf v\cdot r}\rangle&=&0,\nonumber\\
\frac{d\langle {\bf v\cdot r}\rangle}{dt}-\langle {\bf v}^2\rangle+\omega_0^2\langle {\bf r}^2\rangle&=&0,\nonumber\\
\frac{d\langle {\bf v}^2\rangle}{dt}+2\omega_0^2\langle {\bf v}\cdot{\bf r}\rangle&=&0.\label{momentmono}
\end{eqnarray}
Utilizing a trial solution of the form $f({\bf r},{\bf v},t)=e^{i\omega t}f({\bf r},{\bf v})$, allows Eq.~(\ref{momentmono}) to be solved algebraically.  The solution confirms Boltzmann's general result as previously discussed in Sec.~II  for an undamped monopole mode at oscillation frequency $2\omega_0$.  

One can also take $(l=2,m=0)$ quadrupole moments of Eq.~(\ref{momenteq}) to derive a closed set of six coupled equations for an isotropic trap.  However, for the quadrupole mode the moment, $\langle 3{v^2_z}-\bf{v}^2\rangle$ cannot be expressed in the form of Eq.~(\ref{qcons}), which leads to a nontrivial collisional contribution $\langle(3{v^2_z}-{\bf v}^2)I_{coll}\rangle$ in its equation of motion.  For small amplitude oscillations about equilibrium, collisional contributions of this form can be approximately rewritten in the relaxation time approximation \cite{huang,guery1999,pethick} as
\begin{equation}
\langle \chi I_{coll}\rangle=-\frac{\chi}{\tau_{coll}},\label{relax}
\end{equation}
where the relaxation time $\tau_{coll}$ is given by 
\begin{equation}
\tau_{coll}=\frac{5}{4\gamma_{coll}},\label{tau}
\end{equation}
and $\gamma_{coll}=n(0)v_{th}\sigma/2$ is the average collision rate with $v_{th}=\sqrt{8k_b T/\pi m}$.  As with the monopole mode, the set of coupled equations for the quadrupole mode can also be solved algebraically, giving frequencies of $2\omega_0$ and $\sqrt{2}\omega_0$  in the collisionless and hydrodynamic regimes respectively.  However, in the hydrodynamic crossover the quadrupole mode damps as in Eq.~(\ref{relaxation}) due to the nonvanishing collisional contribution.  

\subsection{Scaling ansatz}
The scaling ansatz method has been used for Bose gases above and below the transition temperature ~{\cite{scalingblock,mf,mfdiss,shapeoscill}}.  In the nondegenerate regime, the ansatz method has been used to evaluate the effects of the mean field interaction \cite{mf,mfdiss} and to estimate the effects of anharmonic corrections to the trapping potential on the collective mode frequencies and damping rates \cite{shapeoscill}. Here this method is outlined.   

When the thermal cloud is in statistical equilibrium under harmonic confinement, the Boltzmann equation for the equilibrium distribution $f_0({\bf r},{\bf v})$ is just the Poisson bracket:
\begin{equation}
\left(\sum_i^3 v_i\frac{\partial f_0}{\partial r_i}-\omega_i^2r_i\frac{\partial f_0}{\partial v_i}\right)=0,\label{eqboltz}
\end{equation} 
where the subscript denotes components of a Cartesian vector in 3D and the possibly differing frequency components in each direction.  Moments of the equilibrium distribution are the same as those obtained in the previous subsection when the temporal derivatives are set to zero.  

To describe time-dependent collective oscillations, a scaling ansatz on the form of the distribution function $f({\bf r},{\bf v},t)$ can be made:
\begin{equation}
f({\bf r},{\bf v},t)=\Gamma f_0({\bf R({\bf r},t)},{\bf V({\bf r},{\bf v},t)}).\label{scalingansatz}
\end{equation}
This ansatz utilizes the symmetry of the problem and the nature of the collective modes in the forms of the renormalized position, ${\bf R({\bf r},t)}$, and velocity, ${\bf V({\bf r},{\bf v},t)}$, with an additional factor $\Gamma$ to enforce the normalization.  

For the dipole mode, the form of the ansatz is $R_i=r_i-\eta_i(t)$ and $V_i=v_i-\eta_i(t)$ with $\Gamma$ equal to unity.  The details of the motion are contained in the time-dependent vector of free parameters, $\vec{\eta}$.  Such a scaling clearly describes a translation of the COM of the equilibrium cloud.

 For the monopole and quadrupole mode, the ansatz mimics the form of Boltzmann's solution as discussed in Sec.~II:  
 \begin{eqnarray}
 R_i&=&\frac{r_i}{b_i(t)},\nonumber\\
 V_i&=&\frac{1}{\theta_i^{1/2}}\left(v_i-\frac{\dot{b}_i(t)}{b_i(t)}r_i\right),\nonumber\\
 \Gamma&=&\frac{1}{\prod_j^3(b_j(t)\theta_j(t)^{1/2}),}\label{breathansatz}
 \end{eqnarray}
 where there are two time-dependent vectors of free parameters $\vec{\theta}$ and $\vec{b}$.  Such a scaling describes the stretching and compression of the equilibrium cloud in phase space along with a space-dependent and time-dependent translation in the local velocity, while maintaining a stationary COM. 
 
Substituting Eq.~(\ref{scalingansatz}) into Eq.~(\ref{boltz}) and following \cite{mf,mfdiss}, a Newton-like set of equations of motion for the free parameters can be derived
\begin{eqnarray}
\ddot{b}_i+\omega_i^2b_i-\omega_i^2\frac{\theta_i}{b_i}=0\nonumber\\
\dot{\theta}_i+2\frac{\dot{b}_i}{b_i}\theta_i=-\frac{1}{\tau_{coll}}\left[\theta_i-\bar{\theta}\right],\label{newton}
\end{eqnarray}
where the quantity $\bar{\theta}=\sum_i\theta_i/3$ is the average temperature.  To obtain information about the collective oscillations, the free parameters are linearized about the equilibrium position for  small amplitude oscillation, $b_i\approx1$ and $\theta_i\approx1$, and assumed to have time-dependence of the form $e^{i\omega t}$.  
For a harmonic trap, the results obtained agree with the moment method.

\subsection{Modeling trap imperfections}
  Slight anisotropies in the harmonic confinement and anharmonic corrections to the trapping potential Eq.~(\ref{potential}) present possible sources of damping.  Modeling these imperfections is discussed in this subsection using trapping data from the JILA experiment.  
\subsubsection{Damping due to trap anisotropies}
Anisotropies in the confining potential lead to coupling between the monopole and quadrupole modes, which are damped in the hydrodynamic crossover regime.  The moment method can account for this coupling in a straightforward way given a potential of the form
\begin{equation}
U(x,y,z)=\frac{m\omega_0^2}{2}\left((1+\epsilon)x^2+(1-\epsilon)y^2+\lambda^2 z^2\right),\label{impotential}
\end{equation}
where $\lambda=\omega_z/\omega_0$, $(1+\epsilon)\omega_0=\omega_x$, and $(1-\epsilon)\omega_0=\omega_y$.  

Starting with $\langle {\bf r}^2\rangle$, the moment method yields a set of nine coupled equations for the monopole and quadrupole modes (see Ref.~\cite{spinup}).   In Fig.~1, frequency measurements from the JILA experiment with various anisotropies were used to estimate the damping rate for small oscillations.  For nonzero $\epsilon$ and $\lambda$, the monopole couples to multiple quadrupole modes and the shape of the curve departs from that expected for the typical transition region shape when $\epsilon=0$.    
\begin{figure}
\begin{center}
\includegraphics[width=3.375in]{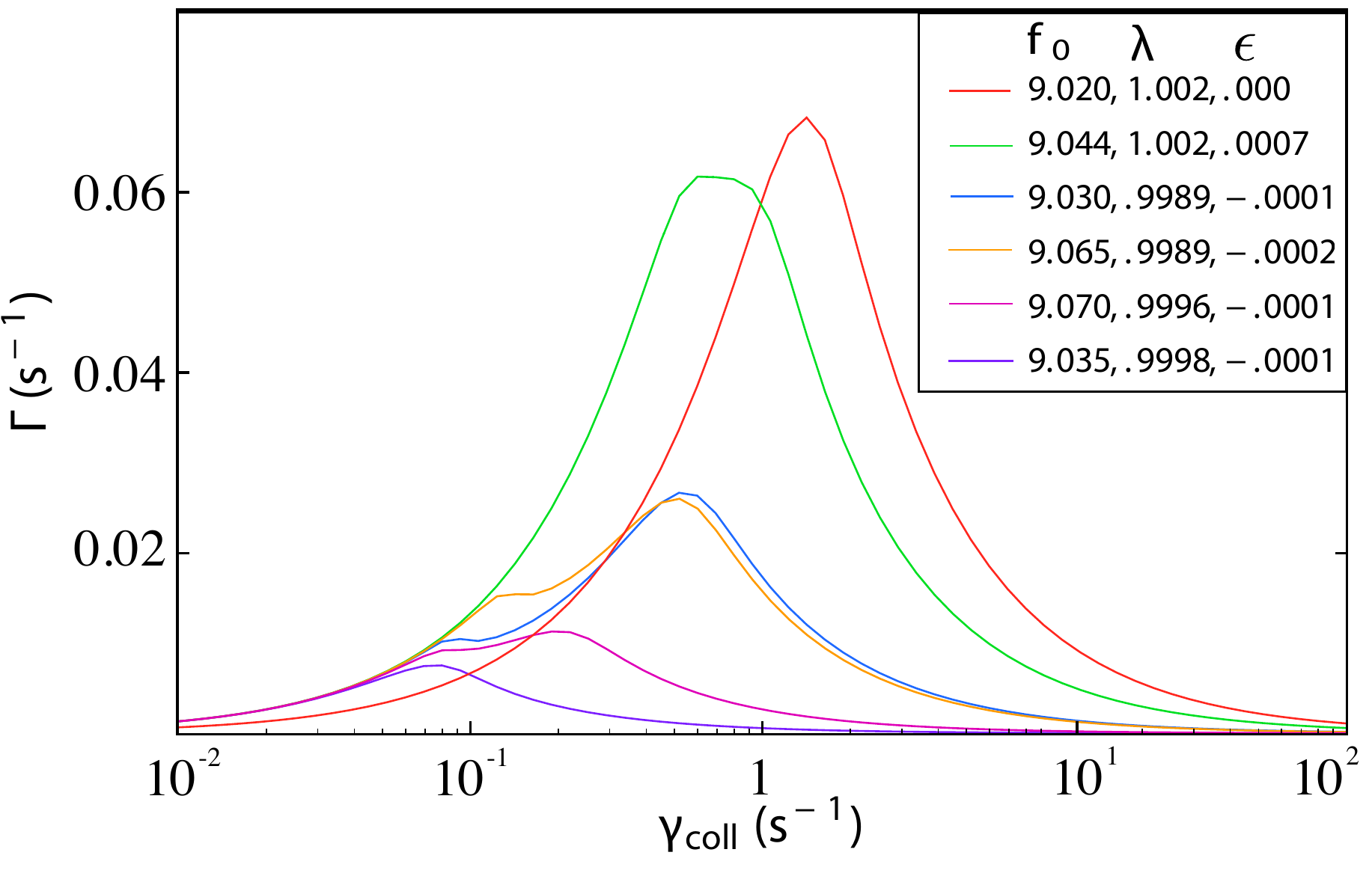}
\caption{Monopole damping rates versus collision rate for a set of frequencies from the JILA experiment with various anisotropy parameters and central frequencies $f_0$.  The legend gives the level of anisotropy, where $\lambda$, $\epsilon$, and $\omega_0$ correspond to rewriting the harmonic confining potential as in Eq.~(\ref{impotential}).}
\end{center}
\end{figure}
The monopole damping measured in the JILA experiment is on the order of $0.1$ s$^{-1}$ or larger.  Therefore, the trap can roughly be treated as isotropic with $\lambda=1$ and $\epsilon=0$ for the lowest four curves in Fig.~1.  Thus, the observed monopole damping must then be due to some other mechanism that is present even when the harmonic piece of the trapping potential is virtually isotropic.

\subsubsection{Damping due to anharmonic corrections}

Computing the monopole damping using the moment method for nonzero anharmonic coefficients generates an infinite set of coupled equations for successively higher order moments and so is not a viable method.  However, the scaling ansatz method, which takes as an input moments of $f_0$, can provide an estimate for the effects of the anharmonic coefficients which enter through higher order moments of $f_0$.  The usefulness of such an approach hinges on being able to neglect deformations of the equilibrium distribution due to the anharmonic corrections.    

Using the scaling ansatz, Eq.~(\ref{newton}) can be rewritten in a more general form for an arbitrary trapping force, $F_i$:
\begin{eqnarray}
\ddot{b}_i-\frac{\theta_i}{b_i}\langle v_i^2\rangle_0-\frac{1}{m}\langle F_i(b_jr_j)r_i\rangle_0=0,\nonumber\\
\dot{\theta_i}+2\frac{\dot{b}_i}{b_i}\theta_i=-\frac{1}{\tau_{coll}}\left[\theta_i-\bar{\theta}\right],\label{newtongen}
\end{eqnarray}
where the 0 subscript indicates a moment as in Eq.~(\ref{avgmoment}) over the equilibrium distribution $f_0$, and $b_j r_j$ is a short hand notation for the set $\{b_xr_x,b_yr_y,b_zr_z\}$.  From Eq.~(\ref{newtongen}), the procedure for including even order corrections to the trapping potential is straightforward.  However, to include odd order corrections, the equilibrium distribution, $f_0$, must be deformed.  A simple deformation is
\begin{equation}
f_0^{ah}({\bf r},{\bf v})=\left(1-\frac{m\beta}{3k_b T}\left(x^2+y^2\right)z-\frac{m\alpha}{3k_b T}z^3\right)f_0({\bf r},{\bf v}),\label{modgauss}
\end{equation}  
which is a perturbative expansion to first order in the odd order anharmonic corrections from Eq.~(\ref{potential}). 
  The deformation does not change the overall normalization nor the value of the even-order moments, and the added terms are collisional invariants of the form Eq.~(\ref{qcons}).   

In Sec.~VI, the validity of this deformation compared to numerical and experimental results for the damping of the monopole mode is addressed.  However, in the remainder of this Section, the general solution of Eq.~(\ref{newtongen}) is discussed for small amplitude oscillations, following the derivation in Ref.~{\cite{shapeoscill}} and noting that the derivation is the same for $f_0^{ah}$.  

Eq.~(\ref{newtongen}) is solved in both the collisionless and hydrodynamic regimes.  Making the assumption that the monopole damping is much less than the collective mode frequency $(\Gamma_M\ll \omega_M)$ leads to a generally applicable formula Eq.~(\ref{transitionlimit}) for the damping.  In the JILA experiment, $\Gamma_M/\omega_M$ is on the order of $10^{-3}$ to $10^{-2}$.

In the collisionless regime, $\tau_{coll}\rightarrow\infty$, which gives a power law relation between the two scaling parameters: $\theta_i=\frac{1}{b_i^2}$.  Using this power law relation, Eq.~(\ref{newtongen}) collapses into three differential equations for each component of the parameter $b_i$.  For small amplitude oscillations $b_i\approx 1$, and the substitution $\eta_i=b_i-1$ linearizes the three differential equations for each component of $\eta_i$:  
\begin{equation}
\ddot{\eta}_i+\frac{3}{m}\frac{\langle U_i r_i\rangle_0}{\langle r_i^2\rangle_0}\eta_i+\frac{1}{m}\sum_j\frac{\langle U_{ij}r_i r_j\rangle_0}{\langle r_i^2\rangle_0}\eta_j=0,\label{newtoncoll}
\end{equation}
where $U_{i}=\partial U/\partial r_i$ and $U_{ij}=\partial^2 U/\partial r_i\partial r_j$.  Eq.~(\ref{newtoncoll}) can be treated with matrix methods as in Ref.~{\cite{shapeoscill}} to extract the collective mode frequencies.  

In the hydrodynamic regime, $\tau_{coll}\rightarrow 0$, and thus the system is in local equilibrium everywhere.  Furthermore, the components of the $\theta_i$'s must be equal to to their average, $\theta_i=\bar{\theta}$.  From the normalization of the ansatz then follows a power law relation between the parameters:  $\theta_i=1/\prod_j b_j^{2/3}$.  This implies that Eq.~(\ref{newtongen}) collapses to a set of three differential equations for each component of the parameter $b_i$ that couple when the power law relation for $\theta_i$ is substituted.  Making the small amplitude assumption leads to the following equation:
\begin{equation}
\ddot{\eta_i}+\frac{1}{m}\frac{\langle U_i r_i\rangle_0}{\langle r_i^2\rangle_0}\left(\frac{5}{3}\eta_i+\frac{2}{3}\sum_{i\neq j}\eta_j\right)+\frac{1}{m}\sum_j\frac{\langle U_{ij} r_i r_j \rangle_0 }{\langle r_i^2\rangle_0}.  \label{newtonhyd}
\end{equation}
As in the collisionless case, matrix methods can be used to extract the collective mode frequencies for the monopole mode and the (l=2,m=0) and (l=2,m=2) quadrupole modes.

The calculated mode frequencies have temperature dependence through higher-order moments coming from the anharmonic corrections.  For a given temperature there is a shift in the frequency and damping from that anticipated by the harmonic limit.  In the collisionless limit, this shift in frequency gives rise to dephasing-induced damping and creates the appearance of actual relaxation in the system over short time scales.  As discussed in \cite{shapeoscill}, it is expected that in the collisionless regime the dephasing induced damping should scale proportionately to the shift in the mode frequency, $\Delta\omega\approx\Delta\Gamma$.  In general, it is less clear how $\Delta\omega$ correlates with the shift in the damping rate.  Therefore a transition regime exists where dephasing effects in the cloud are destroyed through collisions, and collisional damping becomes the dominant effect.  Such an effect can account for anomalous damping of the monopole mode in regimes where $\omega_M\tau_{coll}\gg1$, whereas collisional damping dominates when $\omega_M\tau_{coll}\approx1$.  We refer to this regime as the {\it dephasing crossover}, and in Sec.~V this crossover is studied in detail with the aid of numerical simulation.

\section{\label{sec:level4}Numerical Methods}

In this section the numerical algorithm used to simulate the dynamical evolution of the thermal cloud is described.  Numerical simulation of the system isolates the roles of various damping mechanisms and permits quantification of the dephasing effects in the cloud.  The thermal cloud algorithm from Ref.~\cite{ZNG} is adopted, working in the quantum collisional regime \cite{blak} where two-body collisions are s-wave.  However, the many-particle statistics are still classical and well-described by the Boltzmann equation. 
The simulation consists of a swarm of tracer particles that act as a coarse-grained distribution function \cite{testparticle}:
\begin{equation}
f({\bf r},{\bf p},t)\approx  \frac{N_{th}}{N_{tp}} h^3\sum_i^{N_{tp}}\delta({\bf r}-{\bf r_i})\delta({\bf p}-{\bf p_i}),\label{tracerparticle}
\end{equation}
where $N_{th}/N_{tp}$ is a weighting factor.  The sum is over the entire set of tracer particles, where each is uniquely described by their position and momentum $({\bf r_i}, {\bf p_i})$.  The tracer particles first undergo collisionless evolution via a second order sympletic integrator \cite{splitstep,splitstep2}.  Following free evolution, the tracer particles are binned in space and tested for collisions.  

\subsection{Collisions}

The collision algorithm, which is a procedure for Monte Carlo sampling the collision integral \cite{numrecipes}, follows that described in Ref.~\cite{ZNG}.  After binning the particles in space, random pairs in each bin are selected and their collision probability is calculated.  If a collision is successful, the particle velocities are updated according to the differential cross-section for an s-wave collision. 

For each set of simulation parameters we check that the equilibrium collision rate averaged over the entire cloud matches with the analytic result: 
\begin{equation}
\gamma_{eq}=N_{th}\frac{ \sigma\omega_0^3 m}{2\pi^2 k_b T},\label{eqcoll}
\end{equation}
to  $< 2\%$.  The velocity independent cross-section used is that for bosons $\sigma=8\pi a^2$.

\subsection{Drive mechanism}
To mimic the experimental monopole drive scheme, the trap frequency, $\omega_0$, is modulated over four periods of monopole oscillation
\begin{equation}
\omega_0(t)=\left\{
\begin{array}{lr}
      \omega_0\left(1+A \sin(2\omega_0 t)\right)  &   t\leq 4\pi/\omega_0\\
       \omega_0 & t>4\pi/\omega_0\\
\end{array}
\right\},\label{drive}\\
\end{equation}
\begin{figure}
\begin{center}
\includegraphics[width=3.375in]{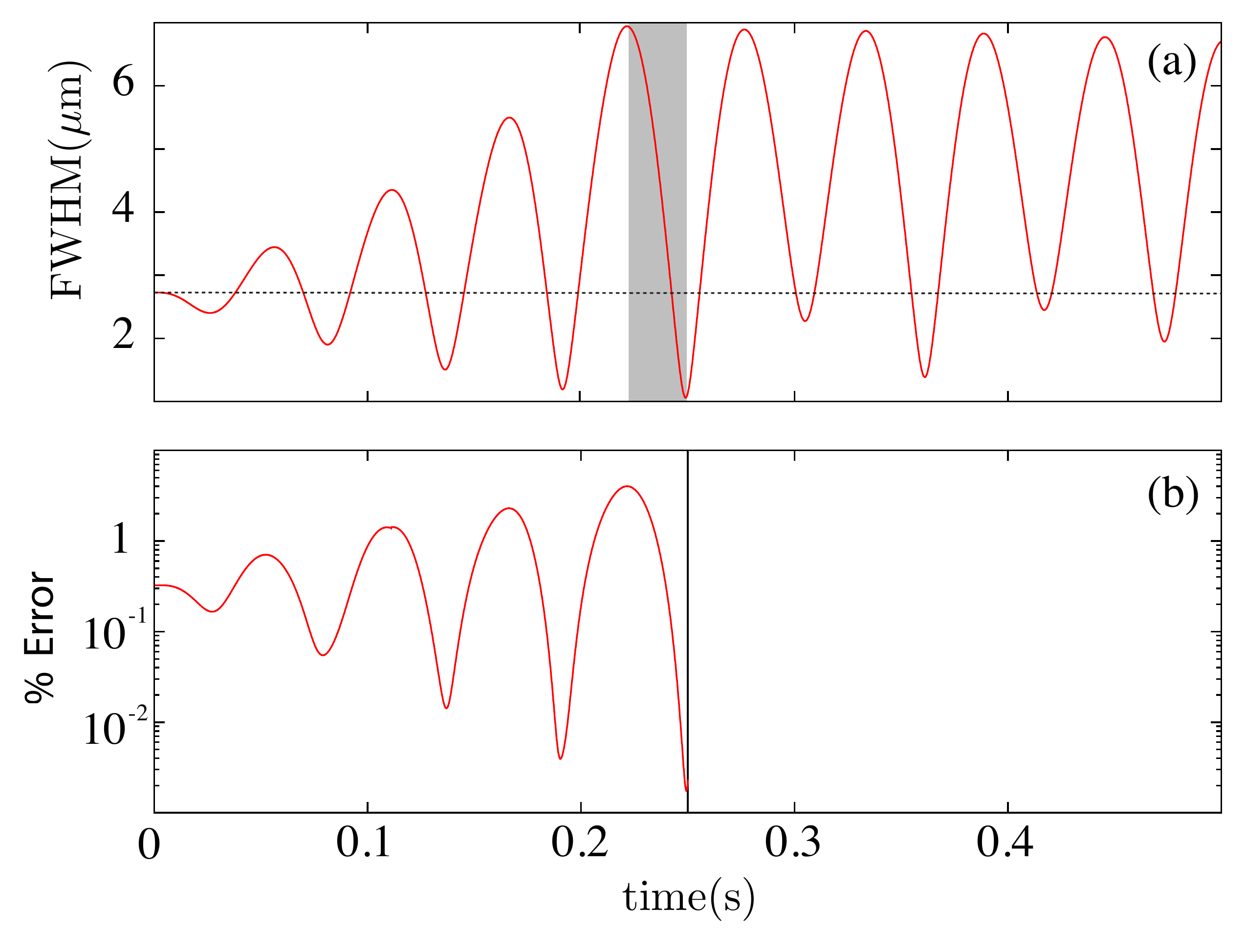}
\caption{Driving the monopole moment for $A=0.15$ and initial temperature of 152nK over four periods of oscillation.  (a)  FWHM of the monopole moment versus holding time.  The grey area indicates the time in between the end of the drive and switching on the anharmonic terms in the trapping potential.  (b) The $\%$ change in the total energy of the cloud between a trapping potential that is purely harmonic and the same potential with anharmonic terms switched on.  The vertical line indicates the point at which the anharmonic terms are added to the trapping potential when the $\%$ increase of the total energy is on the order of $0.001\%$.}
\end{center}
\end{figure}
where $A$ is a unitless measure of the strength of the drive.  During the drive, the anharmonic corrections are neglected, and are switched on when the next oscillation minimum occurs after the drive is turned off.  Fig.~2 illustrates the increase of the monopole amplitude during the drive and the effect of turning on the anharmonic shifts on the total cloud energy.  For $A=0.15$ and a starting temperature of 152nK, the percent increase in the total energy when the anharmonic corrections are switched on is on the order of $0.001\%$.  During the drive there is also a shift in the average temperature and full width at half maximum (FWHM) of the oscillation due to an increase in the energy of the cloud from the work done on it by pumping of the trap.  It would be prohibitive to include the modulation of the bias, quadrupole, and all of the shimming fields used experimentally to recreate the drive.  The simulation driving scheme effectively captures the experimental drive.  Results are presented then over a range of drive strengths which produce an increase in the mean cloud side at the end of the drive phase which is on the order with that experimentally measured.

\section{\label{sec:level5}Results}

We are now in a position to compare and contrast the results of the JILA experiment against the theoretical model and numerical simulation.  Here, only data with the monopole drive is considered.  Moreover, our analysis is focused on experimental data that is not dominated by effects due to trap anisotropy, as this is well-understood \cite{guery1999,pethick}. Characterizing the sensitivity of the monopole mode to anharmonic corrections around the dephasing crossover is the main result of this paper.  This sensitivity stands as a general issue for undamped, nonequilibrium collective modes, such as the monopole oscillation.  The quadrupole modes must also damp around the dephasing crossover; however, these modes are not explored further as they do not fall into the category of undamped, nonequilibrium modes.  

In the JILA experiment, the monopole data was taken for a range of atom numbers and temperatures.  The small cloud data ($N\approx 10^4$ atoms) which lies in the collisionless regime is analyzed first.  Finally the large cloud data ($N\approx 10^5-10^6$ atoms) which lies between the collisionless and hydrodynamic regimes is analyzed.   

\subsection{Collisionless regime}  
In the collisionless regime, the many-body dynamics of the thermal cloud are dictated by the single-particle trajectories.  Here, the system behaves in analogy with a simple pendulum.  For small amplitude oscillations the pendulum executes simple-harmonic motion.  However, as the amplitude grows, the small-angle approximation breaks down and anharmonic corrections become important.  The pendulum traces out a repeating trajectory but with an energy-dependent period.    

The procedure for replacing anharmonic corrections by an energy-dependent harmonic potential in 1D is well-known \cite{landau}.  The renormalized trapping potential for small amplitude motion along the z-axis is:
\begin{eqnarray}
U(0,0,z)&\approx&\frac{m \tilde{\omega}_z^2(E)}{2}z^2,\nonumber\\
\tilde{\omega}_z&=&\omega_z(1+\xi_z E),\nonumber\\
\xi_z&=&-\frac{5\alpha^2}{6m^3\omega_0^6}+\frac{3\kappa}{4m^2\omega_0^4}.\label{renormpot}
\end{eqnarray}
In Fig.~3(a), atoms in the anharmonic trap have been binned according to their energy and the collective motion in each bin averaged to obtain the monopole period in each bin.  For the first several bins the period of collective oscillation increases linearly with the bin energy.  For higher energy bins the scaling is still monotonic; however, the deviation increases as the difference in the linear correction between the different axes becomes more apparent.  At sufficiently high binning energy, the linear correction Eq.~(\ref{renormpot}) breaks down as higher-order terms become important, and the atom number in each bin decreases, spoiling the appearance of an undamped collective mode.    

\begin{figure}[t]
\begin{center}
\includegraphics[width=3.375in]{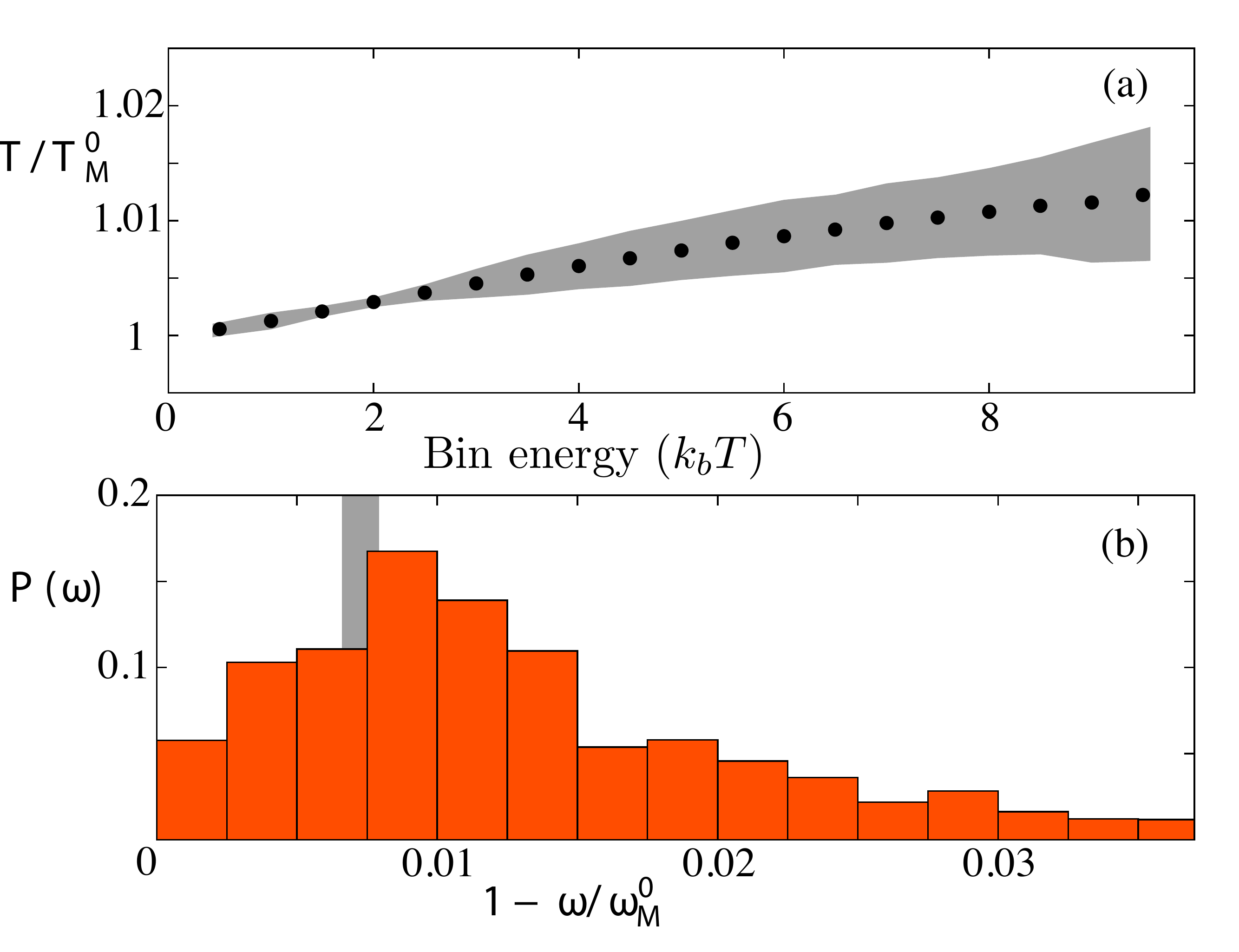}
\caption{Resolving the energy dependence of the monopole moment for a drive strength of $A=0.15$ ($9\pm1.35 $Hz) and an initial temperature of 152nK.  (a) After binning the tracer particles into energy bins, each bin oscillates as an independent monopole.  Here, the ratio of the average period of the monopole in each bin and the zero temperature result, $T^0_M=2\pi/\omega_M^0$ along with the standard deviation (grey region), is plotted versus the energy of each bin in units of thermal energy. (b) The statistical weight, $P(\omega)$, of each frequency component in the cloud.  The grey region is the theoretical prediction for the shift of the monopole frequency from L (150nK) to R (180nK).}
\end{center}
\end{figure}
  
 \begin{figure}[t]
\begin{center} 
\includegraphics[width=3.375in]{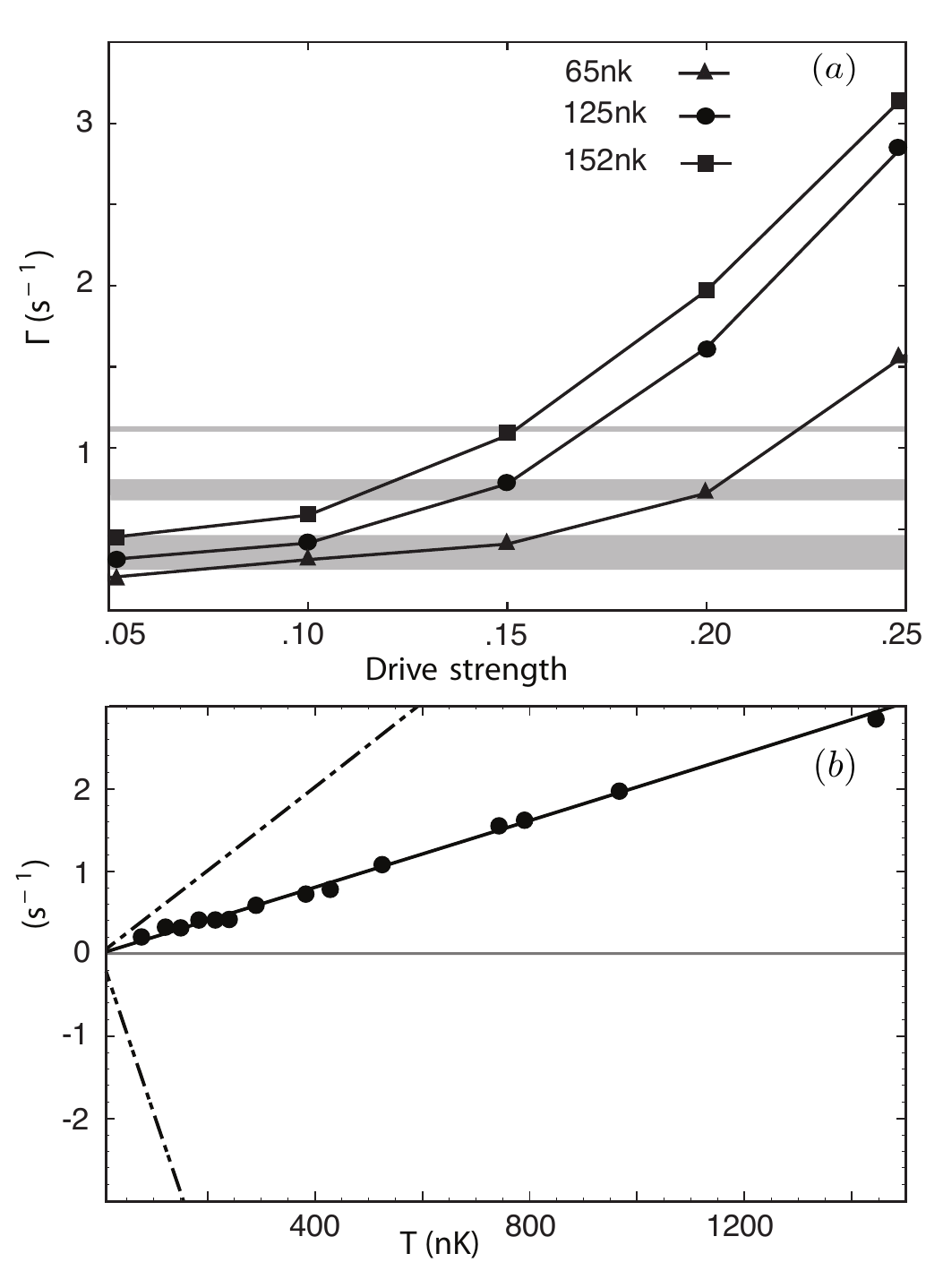}
\caption{(a) Damping of the monopole over a range of drive strengths $A=\{0.05,0.1,0.15,0.20,0.25\}$ and temperatures in the collisionless regime.  The grey regions are the damping results along with the uncertainty from the JILA experiment for (from bottom to top) 65nK, 125nK, and 152nK.  (b)  The dashed lines are the theoretical predictions for $\Delta\omega_M(T)$ using $f_0^{ah}$ (single dash) and using $f_0$ (double dash) for the scaling ansatz versus the settled temperature of the cloud.  The points are the damping results from (a).  The solid line is a fit to the numerical damping results using Eq.~(\ref{scaling}).  }
\end{center}
\end{figure}
  
In the absence of collisions, the population of atoms in each bin is static over the entire simulation.  The statistical weight of each frequency component, as shown in the normalized histogram of Fig.~3(b), is then also static.  The determining characteristics of the frequency distribution are its width and shift of the peak from $\omega_M^0=2\omega_0$.  The theoretical prediction of $\Delta\omega_M$ from the scaling ansatz using $f_0^{ah}$ is for the shift of the peak (grey region of Fig.~3(b)) and agrees with the numerical result.  However, the width, denoted $\delta\omega_M$, of the frequency distribution determines the relative importance of dephasing effects.  If the frequency distribution is shifted and sharply peaked, dephasing induced damping is minimal, mimicking the large amplitude oscillation of a single pendulum.  Otherwise, if the distribution is broad, as in Fig.~3(a), the net sum of many `pendulums' oscillating at shifted frequencies is dephasing induced damping through interference.
 
 We now compare directly to the small cloud data from the JILA experiment.  Starting with the temperature and atom number quoted in the experiment, the drive strength is varied and the amplitude of the resulting oscillation is fit to a simple exponential decay to compare directly to the fitting function used in the experiment.  As shown in Fig.~4(a), a drive strength of $A=0.15$ matches the experimental data, suggesting that the anomalous damping seen in the experiment for the small cloud data is mainly due to dephasing effects.  Such effects depend on the settled temperature of the cloud as shown in Fig.~4(b).

 Although the width of the frequency distribution determines the strength of the dephasing induced damping, in the collisionless regime where the distribution is broad, one naively expects that the frequency shift $\Delta\omega_M$ and the width scale proportionately and can be used as a fitting function, looking then for a scaling fit of the form
 \begin{equation}
 \Gamma_{dephase}(T)=\Delta\omega_M(\zeta T),\label{scaling}
 \end{equation}
where $\zeta$ is a free parameter independent of the temperature.  For the experimental trap parameters, $\zeta=0.4$ from a fit to the data shown in Fig.~4(b).  

The scaling ansatz result using $f_0$ instead of $f_0^{ah}$ predicts a decrease of the period with increasing bin energy, which disagrees directly with the numerical result and Eq.~(\ref{renormpot}).  Additionally, from the dashed lines in Fig.~4(b), as the settled temperature increases the deformed gaussian predicts that the peak frequency decreases ($\Delta\omega_M(T)>0$) as shown by the dashed lines in Fig.~4(b).  
\subsection{Crossover regime}
 The collisionless regime picture of a collection of uncoupled monopole modes oscillating at shifted frequencies begins to break down when the dephasing period $\tau_{dph}=2\pi/\delta\omega_M$
exceeds $\tau_{coll}$.  This is typically in a different collisional regime than the hydrodynamic crossover where $\omega_0\tau_{coll}\approx 1$, and an average atom suffers a collision on a timescale faster than the oscillation period.  Therefore, as the collision rate increases there are two important regimes: the dephasing crossover and the hydrodynamic crossover.  
\begin{eqnarray}
\delta\omega_M\tau_{coll}&\approx& 1 \text{   (Dephasing Crossover),}\label{dephasingeqn}\\
\omega_0\tau_{coll}&\approx& 1 \text{  (Hydrodynamic Crossover).}\label{crossSum}
\end{eqnarray}
In the remainder of this subsection an individual experimental run from the large cloud data set at $60.6$nK with $3.463\times10^5$ atoms with measured trap frequencies $(f_x=9.036 \text{Hz},f_y=9.034 \text{Hz},f_z=9.034\text{Hz})$ is analyzed.  From the criteria Eq.~(\ref{crossSum}) can be obtained crude estimates for the dephasing crossover $\gamma_{coll}\approx 10^{-1}-1$ s$^{-1}$ and the hydrodynamic crossover $\gamma_{coll}\approx 10^2$ s$^{-1}$.  This data set corresponds to the lowest curve in Fig.~1 for a quoted collision rate $\gamma_{coll}=8.88$ s$^{-1}$ which lies between the two regimes.  The maximum damping rate due to anisotropies from Fig.~1 is on the order of $10^{-3}$ s$^{-1}$ compared to the experimentally measured rate $0.14\pm0.02$ s$^{-1}$.  Therefore, the harmonic part of the trapping potential is treated as isotropic.   
 
\begin{figure}[t]
\begin{center} 
\includegraphics[width=3.375in]{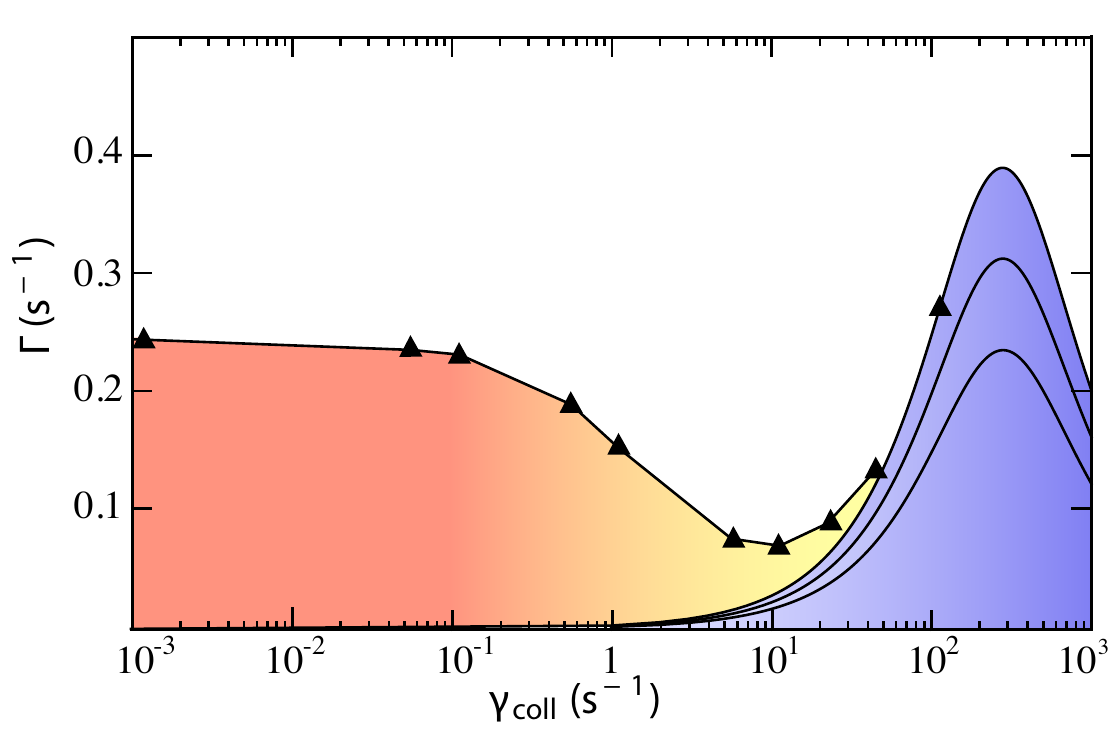}
\caption{Damping rate versus the equilibrium collision rate.  The triangle data points are a result of numerical particle simulations.  The blue curves are scaling ansatz predictions using $f_0^{ah}$ for collisional damping of the monopole mode for temperatures 60nK, 80nK, and 100nK from bottom to top, respectively.  }
\end{center}
\end{figure}

To quantify the effect of the anharmonic corrections in the crossover regime, numerical simulations with $A=0.05$ over a range of collision rates using the trapping frequency $f_0=9.035\text{Hz}$ and initial temperature $60$nK were performed.
Fig.~5 contains the results along with the prediction for the collisional damping from the scaling ansatz theory using $f_0^{ah}$.  The scaling ansatz prediction is plotted for several temperatures, illustrating the effect of the increase in temperature of the cloud post-drive on the damping.  

The estimates for the hydrodynamic and dephasing crossover regimes also agree with the qualitative structure of Fig.~5.  The dephasing crossover is marked by a decrease in the dephasing induced damping; whereas the hydrodynamic crossover is marked by an increase in the collisional damping.  The region in between the two crossovers is characterized by a local minimum in the damping rate.  Comparing the $65$nK damping rate from Fig.~4 with the $60.6$nK damping rate from this subsection draws experimental support for this result.  That the collisionless regime is characterized by higher damping than the dephasing crossover is a common feature of all of the large cloud data compared for similar temperatures to the small data \cite{dan}.  The intermediate regime between dephasing and hydrodynamic crossovers then provides an experimental window where damping from trap anharmonicities can be minimized.  

\begin{figure}[t]
\begin{center}
\includegraphics[width=3.375in]{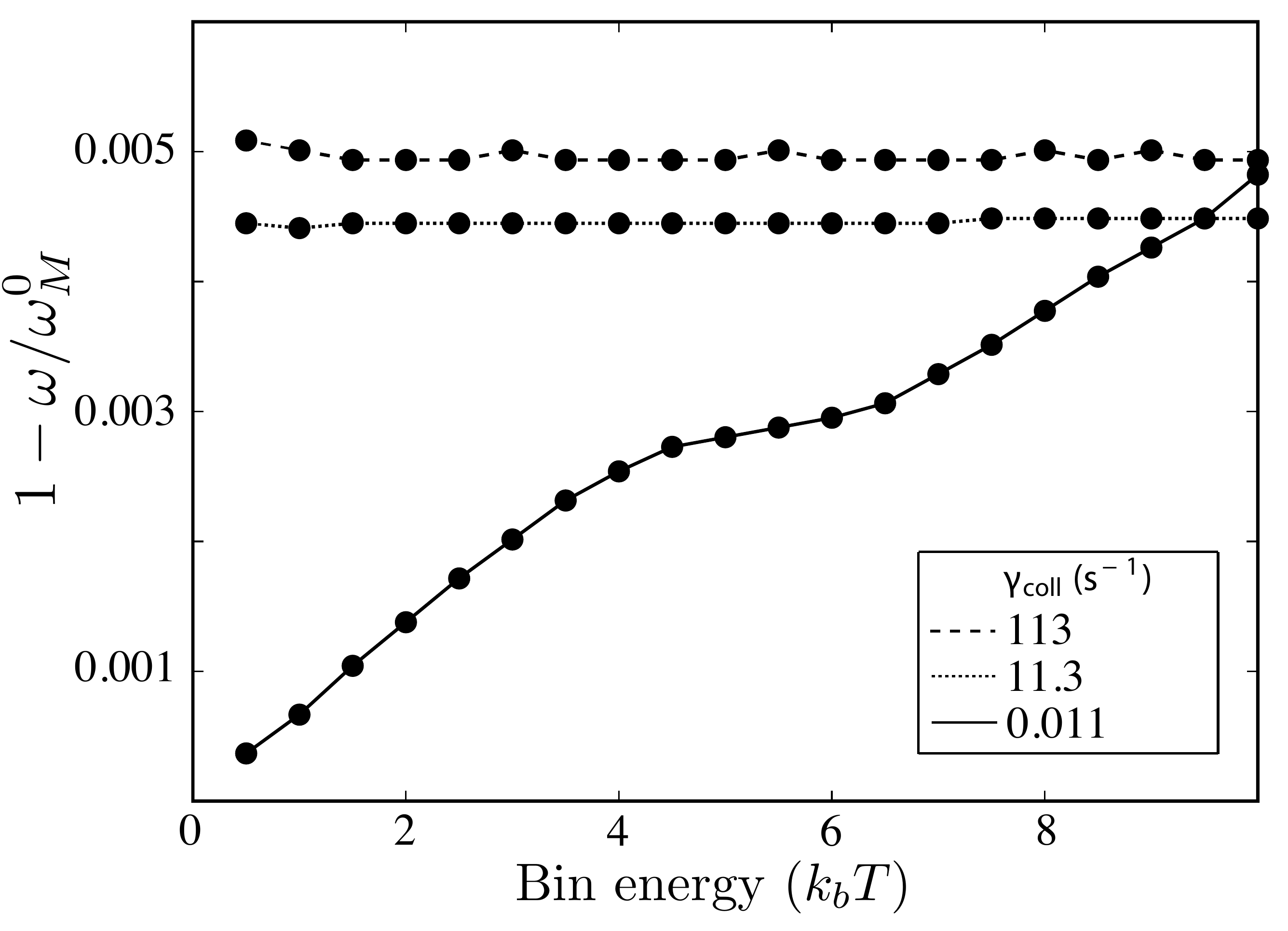}
\caption{Resolving the energy dependence of the monopole moment for A = 0.05 and initial temperature of $60.6$nK for a range of total atom numbers.  Each data point is the peak component of the Fourier transform of the normalized collective oscillation in the bin.  From bottom to top the curves represent frequency spreads in the collisionless, dephasing crossover, and hydrodynamic crossover respectively.}
\end{center}
\end{figure}
Whereas the physics of dephasing in the collisionless regime is analogous to the uncoupled oscillation of a collection of pendulums with different oscillation energies, in the dephasing crossover the pendulums begin to couple, and, as illustrated in Fig.~6, the width of the spread in frequency components narrows.  The coupling effectively synchronizes the different pendulums, which leads to a local minima in the damping rate as seen in Fig.~5.  As the coupling increases, the system nears the hydrodynamic crossover, and the spread in the frequencies is minimal.  The damping is then due to the appearance of a temperature dependent anisotropic trap, where the monopole mode damps naturally through coupling with the quadrupole modes whose moments are not collisionally invariant.  
\section{Conclusion}
In a 3D isotropic trap the monopole mode executes undamped nonequilibrium oscillations as predicted by Boltzmann.  Such a setup has been recently realized experimentally at JILA experiment; however, this experiment measured anomalous damping rates for the monopole mode.  In this paper, possible sources of damping given a realistic trapping scenario based on the JILA experiment are presented.  Slight anisotropies in the trapping frequencies were shown to cause negligible damping.  In contrast, anharmonic corrections to the trapping potential can account for the damping and persist even in the collisionless regime where dephasing effects mimic actual decoherence of the system.  As the collision rate is ramped up, the system traverses the dephasing crossover regime, which is characterized by a local minimum in the damping rate.  A reduction in the damping is consistent with the rates observed experimentally and provides a region of minimum damping where the effects of trap imperfections are reduced.  Quantifying the impact of anharmonic corrections is critical to characterizing damping in the monopole mode in virtually isotropic traps.  

\section{Acknowledgements}

We acknowledge Eric Cornell, Heather Lewandowski, Dan Lobser, and Andrew Barentine for providing experimental data and details.  We also acknowledge helpful discussion with Andrew Sykes.  This material is based upon work supported by the National Science Foundation under Grant Number 1125844 and the Air Force Office of Scientific Research under Grant Number FA9550-14-1-0327.

\end{document}